\documentclass[12pt]{article}
\title{Hyperbolic quantum mechanics}

\author{Andrei Khrennikov\\
International Center for Mathematical\\
Modeling in Physics and Cognitive Sciences,\\
MSI, University of V\"axj\"o, S-35195, Sweden\\
Email:Andrei.Khrennikov@msi.vxu.se}

\begin{document}
\maketitle

\begin{abstract} We start to develop the quantization formalism in a hyperbolic Hilbert
space. Generalizing Born's probability interpretation, we found
that unitary transformations in such a Hilbert space represent a new class of transformations
of probabilities which describe a kind of {\it hyperbolic interference}.
The most interesting problem which was generated by our investigation is to find  experimental evidence of
hyperbolic interference.
The hyperbolic quantum formalism can also be interesting as a new theory of probabilistic
waves that 
can be developed parallely to the standard quantum theory. Comparative analysis
of these two wave theories could be useful for understanding of the role of various structures 
of the standard quantum formalism. In particular, one of 
distinguishing feature of the hyperbolic quantum formalism is the restricted validity
of the superposition principle.
\end{abstract}

\section{Introduction}

\medskip

We develop a formalism that might be called {\it hyperbolic 
quantum formalism.} Instead of the system of complex numbers
${\bf C},$  we use the system of so called hyperbolic numbers\footnote{We remark that
the hyperbolic arithmetics is a straightforward generalization of the complex arithmetics.
Therefore this paper is quite simple from the mathematical viewpoint. We hope that it could
be interesting researches working in theoretical and experimental physics.}
${\bf G}$ (see, for example, [1], p. 21); `physical states'
are represented by vectors in a ${\bf G}$-Hilbert space.
Generalizing Born's probability interpretation, we found
that ${\bf G}$-linear unitary transformations represent a new class of transformations
of probabilities which describe a kind of {\it hyperbolic interference}:
\begin{equation}
\label{LH1}
{\bf P}^\prime ={\bf P}_1 + {\bf P}_2 \pm 2 \sqrt{{\bf P}_1{\bf P}_2} \cosh \theta\;.
\end{equation}
The present formalism is nothing than a theory of {\it hyperbolic waves} of probability
(compare with ${\bf C}$-quantum formalism, a theory of trigonometric waves of probability).

The most interesting problem which was generated by this investigation is to find
hyperbolic interference in experiments with elementary particles (or macro systems). It might
be that such results were already recoded in some experiments with elementary particles.
However, they were not interpreted as an evidence of hyperbolic interference.

On the other hand, the hyperbolic quantum formalism can also be interesting as a new theory of {\it probabilistic
waves} that 
can be developed parallely to the standard quantum theory. Comparative analysis
of these two wave theories could be useful for understanding of the role of various structures 
of the standard quantum formalism, compare with [2]. In particular, we reconsider the role
of complex numbers in quantum theory from the purely probabilistic viewpoint. It seems that complex numbers
were introduced into the quantum formalism to linearize the quantum probabilistic transformation:
\begin{equation}
\label{LT1}
{\bf P}^\prime ={\bf P}_1 + {\bf P}_2 + 2 \sqrt{{\bf P}_1{\bf P}_2} \cos \theta\;.
\end{equation}
The linearization is performed by the ${\bf C}$-representation of real probabilities (\ref{LT1})
on the basis of the formula:
\begin{equation}
\label{LT1a}
A+B + 2\sqrt{AB} \cos \theta= \vert \sqrt{A} + \sqrt{B} e^{i \theta}\vert^2\;.
\end{equation}
In the same way, to linearize  probabilistic transformation (\ref{LH1}) we have to use 
hyperbolic amplitudes:
\begin{equation}
\label{LT1b}
A+B \pm 2\sqrt{AB} \cosh \theta= \vert \sqrt{A} \pm \sqrt{B} e^{j \theta}\vert^2\;.
\end{equation}
Here $j$ is the generator of the algebra ${\bf G}$ of hyperbolic numbers:
$j^2=1.$

Other distinguishing feature of the hyperbolic quantum formalism is the {\it restricted validity
of the superposition principle.} A given state could not be decomposed with respect to an
arbitrary complete system of other states. Of course, as in the complex case, we can always
expend a vector with respect to a basis\footnote{There is some difficulty, because the system
of hyperbolic numbers ${\bf G}$ is only a commutative algebra and not a number field. However,
it is not so important, see [1] on the general theory of Hilbert modules over commutative
and noncommutative algebras.}. However, this operation (which is well defined from the mathematical
point of view) is not always meaningful from the physical point of view. Different representations
are not equivalent in the hyperbolic quantum theory.

The present note is just the first step in the development of the hyperbolic quantum formalism. It would be
interesting to develop this formalism as an alternative to the standard quantum formalism. However, the most
interesting problem is to find the place of hyperbolic waves of probability in experimental physics.

The main ideas on hyperbolic quantum theory were presented in author's talk at the International Conference
"Foundations of Probability and Physics", V\"axj\"o, Sweden-2000 (see also [3]). The author 
had numerous discussions on the origin of the quantum transformation of probabilities
with J. Summhammer (see also [4]). I would like to thank him for these conversations.

\section{ Hyperbolic quantum formalism}

{\bf 1. Hyperbolic algebra.}  A hyperbolic algebra {\bf{G}} is a two dimensional real 
algebra with basis $e_0=1$ and $e_1=j, $ where $j^2=1.$ 
Elements of {\bf{G}} have the form $z=x + j y, \; x, y \in {\bf{R}}.$ 
We have $z_1 + z_2=(x_1+x_2)+j(y_1+y_2)$ and $z_1 z_2=(x_1x_2+y_1y_2)+j(x_1y_2+x_2y_1).$ 
This algebra is commutative. We introduce an involution in {\bf{G}} by setting 
$\bar{z} = x - j y.$ 
We set $|z|^2=z\bar{z}=x^2-y^2.$ 
We remark that  $|z|=\sqrt{x^2-y^2}$ is not well defined for an arbitrary $z\in {{\bf{G}}}.$ 
We set ${{\bf{G}}}_+=
\{z\in{{\bf{G}}}:|z|^2\geq 0\}.$ We remark that ${{\bf{G}}}_+$ 
is the multiplicative semigroup: 
$z_1, z_2 \in {{\bf{G}}}^+ \rightarrow z=z_1 z_2 \in {{\bf{G}}}_+.$ 
It is a consequence of the equality 

$|z_1 z_2|^2=|z_1|^2 |z_2|^2.$

Thus, for $z_1, z_2 \in {{\bf{G}}}_+,$ 
we have $|z_1 z_2|=|z_1||z_2|.$ We introduce 

$e^{j\theta}=\cosh\theta+ j \sinh\theta, \; \theta \in {\bf{R}}.$ 

We remark that 

$e^{j\theta_1} e^{j\theta_2}=e^{j(\theta_1+\theta_2)}, \overline{e^{j\theta}} 
=e^{-j\theta}, |e^{j\theta}|^2= \cosh^2\theta - \sinh^2\theta=1.$

Hence, $z=\pm e^{j\theta}$ always belongs to ${{\bf{G}}}_+.$ 
We also have 

$\cosh\theta=\frac{e^{j\theta}+e^{-j\theta}}{2}, \;\;\sinh\theta=\frac{e^{j\theta}-e^{-j\theta}}{2 j}\;.$

 Let  $|z|^2=x^2-y^2 > 0.$ 
We have 

$z=|z|(\frac{x}{|z|}+j \frac{y}{|z|})= \rm{sign}\; x\; |z|\;(\frac{x {\rm{sign}} x}{|z|} +j\;
\frac{y {\rm{sign}} x}{|z|}).$

As $\frac{x^2}{|z|^2}-\frac{y^2}{|z|^2}=1,$  we can represent $x$ sign $x= \cosh\theta$ 
and $y$ sign $x=\sinh\theta, $ where the phase $\theta$ is unequally defined. 
We can represent each $z\in {{\bf{G}}}_+$ as 

$z = \rm{sign}\; x\;  |z|\; e^{j\theta}\;.$ 

By using this representation we can easily prove that ${{\bf{G}}}_+^*=
\{z\in{{\bf{G}}}_+:|z|^2>0 \}$ 
is the multiplicative group. Here $\frac{1}{z}=\frac{{\rm{sign}} x}{|z|}e^{-j\theta}.$ 
The unit circle in ${{\bf{G}}}$ is defined as $S_1 = \{z\in{{\bf{G}}}:|z|^2=1\}
=\{ z= \pm e^{j \theta}, \theta \in (-\infty, +\infty)\}.$ It is a multiplicative
subgroup of ${\bf G}_+^*.$

{\bf 2. Hyperbolic Hilbert space} is 
${{\bf{G}}}$-linear space (module) ${\bf{E}}$ 
with a ${{\bf{G}}}$-linear
product: a map $(\cdot, \cdot): {\bf{E}}\times {\bf{E}} \to {{\bf{G}}}$ that is 

1) linear with respect to the first argument: 

$ (a z+ b w, u) = a (z,u) + b (w, u), a,b \in {{\bf{G}}}, 
z,w, u \in {\bf{E}};$

2) symmetric: $(z,u)= \overline{(u,z)} ;$

3) nondegenerated: $(z,u)=0$ for all $u \in {\bf{E}}$ iff $z=0.$

{\bf Remark.} If we consider ${\bf{E}}$ as just a ${\bf R}$-linear space, then $(\cdot, \cdot)$ 
is a bilinear form which is not positively defined.  
In particular, in the two dimensional case we have the signature: $(+,-,+,-).$

{\bf 2. Linear space representation of states.} As in the ordinary quantum formalism,
we represent physical states by normalized vectors of the hyperbolic Hilbert space:
$\varphi\in {\bf{E}}$ and $(\varphi, \varphi)=1.$
We shall consider only dichotomic physical variables and quantum states belonging to
the two dimensional Hilbert space. So everywhere below ${\bf{E}}$ denotes the two dimensional
space. Let $a=a_1, a_2$ and $b=b_1, b_2$ be two 
dichotomic physical variables. We represent they by  ${{\bf{G}}}$-linear operators:
$\vert a_1> < a_1\vert + \vert a_2> < a_2\vert$ and $\vert b_1> < b_1\vert + \vert b_2> < b_2\vert,$
where $\{\vert a_i>\}_{i=1,2}$ and  $\{\vert b_i>\}_{i=1,2}$ are two orthonormal bases
in ${\bf{E}}.$

Let $\varphi$ be a state (normalized vector belonging to ${\bf{E}}).$ We can perform 
the following operation (which is well defined from the mathematical point of view). 
We expend the vector $\varphi$ with respect
to the basis $\{\vert b_i>\}_{i=1,2}:$
\begin{equation}
\label{E1}
\varphi = \beta_1\vert b_1>+\beta_2\vert b_2>, 
\end{equation}
where the coefficients (coordinates) $\beta_i$ belong to ${\bf G}.$ 
As the basis $\{\vert b_i>\}_{i=1,2}$ is orthonormal, we get (as in the complex case) that:
\begin{equation}
\label{E2}
|\beta_1|^2+|\beta_2|^2=1\;.
\end{equation}
However, we could not automatically use Born's probabilistic interpretation for
normalized vectors in the hyperbolic Hilbert space: 
it may be that $\beta_i \not \in {\bf G}_+$ (in fact, in the complex case we have
${\bf C}={\bf C}_+).$ We say that a state $\varphi$ is {\it decomposable} with respect 
to the system of states  $\{\vert b_i> \}_{i=1,2}$ if 
\begin{equation}
\label{E3}
\beta_i \in {\bf G}_+ \; .
\end{equation}
In such a case we can use Born's probabilistic interpretation of vectors in 
a hyperbolic Hilbert space: 

Numbers ${\bf q}_i= \vert \beta_i\vert^2, i=1,2,$ are interpreted as probabilities
for values $b=b_i$ for the ${\bf G}$-quantum state $\varphi.$

We now repeat these considerations for each state $\vert b_i>$ by using the basis
$\{\vert a_k>\}_{k=1,2}.$ We suppose that each $\vert b_i>$ is decomposable with respect
to the system of states $\{\vert a_i>\}_{i=1,2}.$ We have:
\begin{equation}
\label{E4}
\vert b_1>=\beta_{11} \vert a_1> + \beta_{12} \vert a_2>,\; \; 
\vert b_2>= \beta_{21} \vert a_1> + \beta_{22} \vert a_2>\;,
\end{equation} 
where the coefficients $\beta_{ik}$ belong to ${\bf G}_+.$   We have automatically:
\begin{equation}
\label{E5}
|\beta_{11}|^2+|\beta_{12}|^2=1, \; \;|\beta_{21}|^2+|\beta_{22}|^2=1\;.
\end{equation}
We can use the probabilistic interpretation of numbers ${\bf p}_{11}=\vert \beta_{11}\vert^2,
{\bf p}_{12}=\vert \beta_{12}\vert^2$ and ${\bf p}_{21}=\vert \beta_{21}\vert^2,
{\bf p}_{22}=\vert \beta_{22}\vert^2.$ 

{\bf Remark.} Let us consider matrices $B=(\beta_{ik})$ and $P=({\bf p}_{ik}).$
As in the complex case, the matrice $B$ is unitary: vectors $u_1= (\beta_{11}, \beta_{12})$
and $u_2=(\beta_{21}, \beta_{22})$ are orthonormal. The matrice $P$ is double
stochastic, namely: ${\bf p}_{11}+ {\bf p}_{12}=1, {\bf p}_{21}+ {\bf p}_{22}=1$
and ${\bf p}_{11}+ {\bf p}_{21}=1, {\bf p}_{12}+ {\bf p}_{22}=1.$

By using the ${\bf G}$-linear space calculation (the change of the basis) we get
$\varphi= \alpha_1 \vert a_1> + \alpha_2 \vert a_2>,$
where $\alpha_1= \beta_1 \beta_{11}+ \beta_2\beta_{21}$ and 
$\alpha_2= \beta_1 \beta_{12}+ \beta_2\beta_{22}.$

We remark that decomposability is not transitive. In principle $\varphi$
may be not decomposable with respect to $\{\vert a_i>\}_{i=1,2},$
despite the decomposability of $\varphi$ with respect to $\{\vert b_i>\}_{i=1,2}$
and the decomposability of the latter system with 
respect to $\{\vert a_i>\}_{i=1,2}.$ 

Suppose that $\varphi$ is decomposable with respect to $\{\vert a_i>\}_{i=1,2}.$
Therefore coefficients ${\bf p}_k= \vert \alpha_k\vert^2$ can be interpreted as
probabilities for $a=a_k$ for the ${\bf G}$-quantum state $\varphi.$

As numbers $\beta_i, \beta_{ik}$ belong to ${\bf G}_+,$ we can uniquely represent them
as

$\beta_i=\pm \sqrt{q_i} e^{j \xi_i}, 
\beta_{ik}=\pm \sqrt{p_{ik}} e^{j \gamma_{ik}}, i, k, =1,2.$ 

We find that
\begin{equation}
\label{E7a}
{\bf p}_1= {\bf q}_1{\bf p}_{11} + {\bf q}_2{\bf p}_{21} + 
2 \epsilon_1 \sqrt{{\bf q}_1{\bf p}_{11}{\bf q}_2{\bf p}_{21}} \cosh \theta_1\;, 
\end{equation}
\begin{equation}
\label{E7b}
{\bf p}_2= {\bf q}_1{\bf p}_{12} + {\bf q}_2{\bf p}_{22} +
2 \epsilon_2 \sqrt{{\bf q}_1{\bf p}_{12}{\bf q}_2{\bf p}_{22}} \cosh \theta_2\;,
\end{equation}
where $\theta_i = \eta+ \gamma_i$  and $\eta= \xi_1- \xi_2,
\gamma_1= \gamma_{11}- \gamma_{21}, \gamma_1= \gamma_{12}- \gamma_{22}$
and $\epsilon_i= \pm.$
To find the right relation between signs of the last terms in equations  (\ref{E7a}),
(\ref{E7b}), we use the normalization condition 
\begin{equation}
\label{E7}
\vert \alpha_1\vert^2 + \vert \alpha_2\vert^2=1
\end{equation}
(which is a consequence of the normalization of $\varphi$ and orthonormality of 
the system $\{\vert a_i>\}_{i=1,2}).$ It is equivalent to the equation:
\begin{equation}
\label{E8}
\sqrt{p_{12}p_{22}} \cosh\theta_2 \pm \sqrt{p_{11}p_{21}} \cosh\theta_2=0.
\end{equation}
Thus we have to choose opposite signs in equations (\ref{E7a}), (\ref{E7b}).
Unitarity of $B$ also inply that $\theta_1 - \theta_2 =0.$
We recall that in the ordinary quantum mechanics we have similar conditions, but
trigonometric functions are used instead of hyperbolic
and phases $\gamma_1$ and $\gamma_2$ are such that $\gamma_1 - \gamma_2= \pi.$

Finally, we get that (unitary) linear transformations in the ${\bf G}$-Hilbert space
(in the domain of  decomposable states) represent the following transformation
of probabilities:
\begin{equation}
\label{E7c}
{\bf p}_1= {\bf q}_1{\bf p}_{11} + {\bf q}_2{\bf p}_{21} \pm 
2  \sqrt{{\bf q}_1{\bf p}_{11}{\bf q}_2{\bf p}_{21}} \cosh \theta\;, 
\end{equation}
\begin{equation}
\label{E7d}
{\bf p}_2= {\bf q}_1{\bf p}_{12} + {\bf q}_2{\bf p}_{22} \mp
2  \sqrt{{\bf q}_1{\bf p}_{12}{\bf q}_2{\bf p}_{22}} \cosh \theta\;.
\end{equation}
This is a kind of hyperbolic interference.\footnote{By changing
hyperbolic functions to trigonometric we obtain the standard quantum interference
of alternatives.}

{\bf Remark.} In fact, we first derived the probabilistic transformation
(\ref{E7c}), (\ref{E7d}) by classifying possible probabilistic transformations
induced by perturbation effects [3]. Then we found the corresponding 
linear space representation starting with the analogy between
(\ref{LT1a}) and (\ref{LT1b}). 

{\bf Remark.} There can be some connection with quantization in Hilbert 
spaces with indefinite metric 
as well as the theory of relativity. However, at the moment  we cannot 
say anything definite. It seems that by using  Lorentz-`rotations'
we can produce hyperbolic interference in a similar way as we produce
the standard trigonometric interference by using ordinary rotations.

{\bf References}

[1]  A. Yu. Khrennikov, {\it Supernalysis.}  (Kluwer Academic Publishers, 
Dordreht/Boston/London, 1999).

[2] P. A. M.  Dirac, {\it The Principles of Quantum Mechanics.}
(Claredon Press, Oxford, 1995); J. von Neumann, {\it Mathematical foundations
of quantum mechanics.} (Princeton Univ. Press, Princeton, N.J., 1955);
E. Schr\"odinger, {\it Philosophy and the Birth of Quantum Mechanics.}
Edited by M. Bitbol, O. Darrigol. (Editions Frontieres, 1992);
J. M. Jauch, {\it Foundations of Quantum Mechanics.} (Addison-Wesley, Reading, Mass., 1968);
A. Peres, {\it Quantum Theory: Concepts and Methods.} (Kluwer Academic Publishers, 1994);
B. d'Espagnat, {\it Conceptual foundations of Quantum Mechanics.} (Perseus Books, Reading,
Mass.,1999); P. Busch, M. Grabowski, P.J. Lahti, {\it Operational Quantum Physics.}
(Springer Verlag, 1995).

[3] A. Yu. Khrennikov, {\it Ensemble fluctuations and the origin of quantum probabilistic
rule.} Report MSI, Vaxjo University, N.90, October, (2000).

[4] J. Summhammer, Int. J. Theor. Physics, {\bf 33}, 171-178 (1994);
Found. Phys. Lett. {\bf 1}, 113 (1988); Phys.Lett., {\bf A136,} 183 (1989).

\end{document}